\documentclass[11pt]{article}
\usepackage{times}
\usepackage{geometry}
\usepackage[utf8]{inputenc}
\usepackage{enumitem,amssymb}
\usepackage{ragged2e}
\newlist{thematic}{itemize}{8}
\setlist[thematic]{label=$\square$}
\usepackage{pifont}
\usepackage{amsmath}
\usepackage[backend=biber, style=verbose, autocite=footnote, 
doi=false,
url=false,
arxiv=false,
isbn=false]{biblatex}            

\AtEveryBibitem{\clearfield{title}}

\usepackage[colorlinks=true, linkcolor=blue, citecolor=blue, urlcolor=blue]{hyperref}

\begin{document}

\raggedright
ESO Expanding Horizon White Paper:
\vspace{.5cm}
\\
{\LARGE  
Uncovering the population of compact binary mergers and their formation pathways with gravitational waves through the Einstein Telescope
}\\ 


\vspace{.5cm}

\textbf{Authors:}\\[2mm]
Manuel Arca-Sedda\textsuperscript{\ref{gssi},\ref{gssinfn},\ref{oar}},
Irina Dvorkin\textsuperscript{\ref{iap},\ref{iuf}},
Gabriele Franciolini\textsuperscript{\ref{unipd},\ref{infnpd},\ref{cern}},
M.~C.~Artale\textsuperscript{\ref{unab}},
M.~Branchesi \textsuperscript{\ref{gssi}, \ref{lngs}},
E. Bortolas\textsuperscript{\ref{oapd}},
M. Colpi \textsuperscript{\ref{occhialini}},
V. De Luca\textsuperscript{\ref{jhu}},
Archisman Ghosh\textsuperscript{\ref{ugent}},
M.~Maggiore\textsuperscript{\ref{DPTunige},\ref{gwsc}},
M. Mapelli\textsuperscript{\ref{heidelberg1}, \ref{heidelberg2}, \ref{unipd}, \ref{infnpd}},
B. Mestichelli \textsuperscript{\ref{gssi}, \ref{lngs}, \ref{heidelberg1}},
M. Mezcua\textsuperscript{\ref{ice},\ref{ieec}},
S. Nissanke\textsuperscript{\ref{nissa}},
L. Paiella\textsuperscript{\ref{gssi}, \ref{lngs}},
A.~Riotto\textsuperscript{\ref{DPTunige},\ref{gwsc}},
F. Santoliquido\textsuperscript{\ref{gssi}, \ref{lngs}}
N. Tamanini\textsuperscript{\ref{tama}},
R. Schneider\textsuperscript{\ref{sap}},
C. Ugolini\textsuperscript{\ref{gssi}, \ref{lngs}, \ref{oar}},
M.~P. Vaccaro\textsuperscript{\ref{heidelberg1}}, 
K. Yakut\textsuperscript{\ref{ky1},\ref{ky2}}

{\scriptsize
\begin{enumerate}[label=\alph*), leftmargin=1.5em, itemsep=0pt, topsep=2pt, parsep=0pt]
\renewcommand{\theenumi}{\alph{enumi}}
\item \label{gssi} Gran Sasso Science Institute, Via F. Crispi 7, I-67100 L'Aquila, Italy

\item \label{gssinfn} INFN, Laboratori Nazionali del Gran Sasso, I-67100  Assergi, Italy 

\item \label{oar} INAF – Osservatorio Astronomico di Roma, Via di Frascati 33, I-00078 Monteporzio Catone, Italy

\item \label{iap} Institut d’Astrophysique de Paris, UMR 7095, CNRS and Sorbonne Université, 98 bis boulevard Arago, 75014 Paris, France

\item \label{iuf} Institut Universitaire de France, Minist\`ere de l’Enseignement Sup\'erieur et de la Recherche, 1 rue Descartes, 75231 Paris Cedex F-05, France

\item \label{unipd} Dipartimento di Fisica e Astronomia `G. Galilei'',  Università degli Studi di Padova, via Marzolo 8, I-35131 Padova, Italy

\item \label{infnpd} INFN, Sezione di Padova, via Marzolo 8, I-35131 Padova, Italy

\item \label{cern} Department of Theoretical Physics, CERN, Esplanade des Particules 1, P.O. Box 1211, Geneva 23, Switzerland

\item \label{unab} Universidad Andres Bello, Facultad de Ciencias Exactas, Departamento de Fisica y Astronomia, Instituto de Astrofisica, Fernandez Concha 700, Las Condes, Santiago RM, Chile

\item \label{lngs} INFN, Laboratori Nazionali del Gran Sasso, I-67100 Assergi, Italy

\item \label{oapd} INAF--osservatorio Astronomico di Padova, vicolo dell'Osservatorio 5, I-35122 Padova, Italy

\item \label{occhialini} Dipartimento di Fisica G. Occhialini, Universit\`a di Milano-Bicocca, Piazza della Scienza 3,
I-20126 Milano, Italy

\item \label{jhu} William H.\ Miller III Department of Physics and Astronomy, Johns Hopkins University, 3400 North Charles Street, Baltimore, Maryland, 21218, USA

\item \label{ugent} Department of Physics and Astronomy, Ghent University, B-9000 Ghent, Belgium

\item \label{DPTunige} D\'epartement de Physique Th\'eorique, Universit\'e de Gen\`eve, 24 quai Ernest Ansermet, 1211 Gen\`eve 4, Switzerland

\item \label{gwsc} Gravitational Wave Science Center (GWSC), Universit\'e de Gen\`eve, CH-1211 Geneva, Switzerland

\item \label{heidelberg1} Institut f\"ur Theoretische Astrophysik, Zentrum f\"ur Astronomie, Universit\"at Heidelberg, Albert Ueberle Str. 2, D-69120 Heidelberg, Germany

\item \label{heidelberg2} Interdiszipli\"ares Zentrum f\"ur Wissenschaftliches Rechnen, Universit\"at Heidelberg, D-69120 Heidelberg, Germany

\item \label{ice} Institute of Space Sciences (ICE, CSIC), Campus UAB, Carrer de Magrans, 08193, Barcelona, Spain

\item \label{ieec} Institut d'Estudis Espacials de Catalunya (IEEC),  Edifici RDIT, Campus UPC, 08860 Castelldefels, Barcelona, Spain

\item \label{nissa} GRAPPA, Anton-Pannekoek Institute for Astronomy, Institute of Physics, Universiteit van
Amsterdam, Amsterdam, The Netherlands

\item \label{tama} Laboratoire des 2 Infinis - Toulouse (L2IT-IN2P3), Universit\`e de Toulouse, CNRS, UPS,
F-31062 Toulouse Cedex 9, France

\item \label{sap} Dipertimento di Fisica, Sapienza Università di Roma, Piazzale A. Moro 2, 00185 Roma, Italy 

\item \label{ky1} Department of Astronomy and Space Sciences, Faculty of Science, Ege University, 35100, {\.I}zmir, Turkey

\item \label{ky2} Institute of Astronomy, The Observatories, Madingley Road, Cambridge CB3 OHA, UK

\end{enumerate}
}


\justifying
\pagebreak

\textbf{
Ground-based gravitational-wave (GW) observatories have transformed our view of compact-object mergers, yet their reach still limits a comprehensive reconstruction of the processes that generate these systems. 
Only next-generation observatories, with order-of-magnitude improvements in sensitivity and access to lower frequencies, will be capable of radically extending this detection horizon. GW observations will make it possible to detect the complete population of binary black hole (BBH) mergers out to redshifts of $z \simeq 100$. This capability will deliver an unprecedented map of merger events across cosmic time and enable precise reconstruction of their mass and spin distributions,  while for several thousand events the signal-to-noise ratio will surpass 100, enabling precision physics of BHs and neutron stars (NSs).
The access to lower frequencies will also open the intermediate-mass window, detecting systems of order $\sim 10^3\,M_\odot$, potentially in coordination with multi-band observations from LISA. At  higher redshifts, where Population~III stars have so far remained beyond reach —even for the James Webb Space Telescope— GW observations by next-generation detectors will routinely provide observations of BH mergers thought to originate from these primordial stellar populations. 
Such measurements are expected to play a central role in clarifying the early assembly of supermassive black holes.
A single detection of a binary BH system at $z \gtrsim 30$, or of a compact object with sub-solar mass and no tidal deformability, would constitute strong evidence for the existence of primordial black holes. Such a discovery would have profound consequences for our understanding of dark matter and the early Universe. Ultimately, the GW observations will become revolutionary for identifying the physical channels responsible for compact binary formation.
}
\linebreak




\textbf{\textit{ Introduction.}} Within the last decade, a series of ground-breaking discoveries improved our knowledge of compact objects and black holes (BHs) at all the scales. The LVK collaboration confirmed the existence of BHs and BH binaries through the detection of their GW emission \autocite{LIGOScientific:2016aoc}, enabling exquisite tests of GR \autocite{LIGOScientific:2016lio}, driving a revolution in stellar BH formation theory by discovering BHs as massive as 200 M$_\odot$ \autocite{LIGOScientific:2025rsn}, and confirming that binary neutron stars (NSs) can merge and trigger GRB and kilonova emission \autocite{LIGOScientific:2017ync}, marking the birth of multi-messenger Astronomy. Detailed observations of the Galactic Center have offered new stringent constraints on the properties of the Milky Way SMBH \autocite{GRAVITY:2023avo}. These results include GR tests from the motion of the so-called S stars \autocite{GRAVITY:2018ofz} and the first ``image" of the gas swirling around it\autocite{EventHorizonTelescope:2022wkp}, which was preceeded by the same picture taken in the nucleus of M87 \autocite{EventHorizonTelescope:2019ggy}. 
Hubble space telescope data of the massive globular cluster Omega Centauri have revealed the presence of an intermediate mass BH (IMBH) there, placing stringent constraints on its mass, of around 8,000 Ms, right in the middle of the uncharted "IMBH mass range" \autocite{2024Natur.631..285H}. 

While representing leaps in the understanding of compact objects and their environments, these discoveries opened key questions that current GW detectors are unable to address, especially concerning the origin of compact binary coalesces (CBCs), IMBHs, and SMBHs. Moreover, current detectors are not sensitive enough to uniquely identify mergers involving primordial BHs (PBHs) or, more in general, sub-solar mergers, or to unveil the gravitational-wave background (GWB) from stationary sources. 
In these regards, next-generation observatories, such as the Einstein Telescope (ET) \autocite{ET:2025xjr} will unlock the detection of GW sources in the stellar- and intermediate-mass regime up to the dawn of the Universe, well before the peak of the star formation rate, potentially determining a new leap in our understanding. In fact, future GW observations are expected to profoundly affect several scientific areas, as briefly summarised in the following lines. 

\textbf{\textit{Astrophysical CBC populations}}. 
Accessing the entire population of BBH mergers up to redshift $z \simeq  100$, as well as binary NS mergers beyond the peak of star formation, would allow us to fully uncover the properties CBCs across cosmic time.

The cosmological population of CBCs is expected to be 
a mixture of binaries formed through two main formation channels: isolated or dynamical \autocite{Zevin:2020gbd,Sedda:2021vjh}. 
{\it i)} Isolated CBCs form only through binary stellar evolution \autocite{Belczynski:2001uc}, and their properties are mostly affected by the cosmological formation history \autocite{Vangioni:2015ofa}, the natal kick of compact objects \autocite{Giacobbo:2019fmo}, and the physics of common-envelope and mass-transfer evolution \autocite{Iorio:2022sgz}.
{\it ii)} Dynamical CBCs form via dynamical interactions in star clusters \autocite{PortegiesZwart:1999nm} and are expected to feature high-eccentricity, isotropic orientation of the spins, and BH masses generally larger than in the case of isolated binaries (i.e. $>30M_\odot$) \autocite{Rodriguez:2016kxx,Sedda:2020wzl}. Remnants retained in their clusters can undergo higher-generation mergers, acquiring typical spins $\sim 0.6-0.8$ possibly clearly distinguishable by the distribution of natal spin for stellar BHs \autocite{Gerosa:2021mno}. 

Understanding the properties of compact objects can constrain the cosmological evolution of their stellar progenitors. Population III stars, the first stars appeared in the Universe, represent a vastly unconstrained class of objects that is expected to produce massive CBCs at redshift $>15$ \autocite{Tanikawa:2020cca,Mestichelli:2024djn}.
Hence, constraining the mass spectrum of CBCs at redshift $z\simeq 15-40$ can help understanding better the first phases of star formation and the birth of the first stars. 
%
Additionally, extending the GW observational band to frequencies as low as 2 Hz will allow to uncover and characterise the properties of IMBHs in the still poorly known mass-range between $10^2-10^5$ M$_\odot$. These objects may form in dense stellar clusters, from the collision of stars \autocite{PortegiesZwart:2002iks} or repeated mergers of stellar BHs \autocite{Miller:2001ez}, or could form in the nucleus of galaxies during their assembly and represent the seeds of SMBHs \autocite{1987ApJ...321..199Q}. 
Combining next-generation ground-based GW observatories with detectors sensitive to lower frequencies, thus to larger IMBH masses, will permit to constrain the IMBH mass spectrum in the whole $10^2-10^5$ M$_\odot$ across cosmic ages \autocite{Valiante:2020zhj}. Space-borne detectors like LISA, for example, can observe binary IMBHs out to a redshift $z\gtrsim 100$ and nearby IMBHs merging with stellar BHs or SMBHs at $z<1$ \autocite{LISA:2024hlh}, whilst detectors operating in the deci-Hz frequency band, like the Lunar Gravitational Wave Antenna (LGWA) \autocite{Ajith:2024mie}, could observe merging IMBHs and IMBH-BH binaries out to a redshift $z \sim 1-10$. Synergetic observations of IMBHs with different detectors across multiple bands will enable us understanding the nature of IMBHs, both as the byproduct of "stellar" processes and as the building blocks of SMBHs. 

High signal-to-noise observations of BHs and NSs will enable constraining single and binary stellar evolution, the presence of gaps (possibly linked to the pair-instability supernovae) and peaks in the mass distribution, orbital eccentricity, and spin precession, providing key insights into the astrophysical environments of the mergers. Only a large number of detections  will enable robust statistical inference of more complex and sophisticated models, addressing the interpretation challenges that are currently unavoidable.

\textbf{\textit{ Primordial BH population.}}
PBHs are a hypothetical population of BHs that could have formed in the early Universe from the collapse of extreme density inhomogeneities during the early phases of cosmological expansion \autocite{Byrnes:2025tji}. While tightly constrained as a dominant component of dark matter\autocite{Carr:2020gox}, a subdominant population of PBHs could provide seeds for high-redshift supermassive black holes, explain small-scale structure anomalies, and enhance signals from other dark matter candidates (e.g. weakly interactive massive particles). PBHs can span a broad mass range, including the stellar-mass regime accessible to ground-based GW detectors. Their mass distribution is model dependent and unconstrained, and depends sensitively on the primordial density perturbation spectrum and on the physics of the early Universe. PBHs pair in binaries mainly through gravitational decoupling from the Hubble expansion, occurring before matter–radiation equality, and merge after long delays, making such events potentially observable at late times. Even accounting for current stringent bounds from microlensing surveys such as OGLE, future GW observations will push the sensitivity to PBH mergers far beyond current limits. In particular, the high-redshift window ($z \gtrsim 30$), where no astrophysical mergers are expected at significant rates, offers a unique discovery potential, allowing direct probes of the primordial Universe and potentially linking PBHs to the seeds of supermassive black holes observed today.
An additional smoking-gun signature accessible with next-generation GW detectors lies in the sub-solar mass window, where the exquisite precision is needed to measure the absence of tidal deformability, providing a distinctive test of the primordial scenario or other exotic compact objects.

%

\textbf{\textit{ Stochastic GW backgrounds from CBCs.}}
The incoherent superposition of GW emission from all astrophysical sources throughout the Universe gives rise to a stochastic gravitational-wave background (SGWB) whose amplitude and spectrum encode the characteristics of the underlying source populations. The SGWB is a powerful probe of astrophysical processes, providing information that complements the study of individually resolved sources, retaining information from mergers which are individually unobservable.
Although most BBH mergers are expected to be individually detectable with next-generation GW sources, this will not be the case for the bulk of the BNS population. BNS signals are intrinsically weaker and remain in band for much longer, leading to significant overlap and a large fraction of subthreshold events. A detection of the SGWB will therefore provide access to information about the BNS population that cannot be obtained from a catalogue of individually resolved events alone.
Even for BBHs, however, a background-based inference of parameters such as the merger rate and mass distribution offers an important cross-check against the estimates derived from individually detected events. Any discrepancy between the two could signal the presence of a faint or distant subpopulation that eludes standard detection pipelines.
This is particularly relevant in scenarios involving PBHs, whose population could extend to redshifts well beyond the instrumental horizon; the properties of such a high-redshift population may only be constrained through a background approach in some regimes.
Finally, astrophysical backgrounds act as a foreground for potential primordial signals. Accurately characterizing their spectral and statistical properties is therefore essential for subtracting their contribution from the observed background and for isolating any cosmological components.

\textbf{\textit{Technology development and  requirements.}} 
This white paper demonstrates that a next-generation observatory, such as the Einstein Telescope, with its unprecedented sensitivity and access to frequencies down to 2 Hz, will represent a unique and groundbreaking facility that fundamentally surpasses the capabilities of current detectors. It will be able to address key questions regarding the population properties, formation channels, and host environments of CBCs across the Universe, while also potentially revealing the existence of primordial BHs of cosmological origin.
In addition, its improved parameter-estimation precision will make it possible to detect populations of sub-solar primordial BHs or other exotic objects. The Einstein Telescope will open access to the realm of IMBHs, enabling observations of systems with masses up to $5\times 10^3$ out to redshifts $2\div 10$, depending on the binary mass ratio. It will also enable exploration of the SGWB produced by astrophysical sources with nearly stellar masses.
The level of understanding attainable for these populations of compact objects cannot be achieved with any electromagnetic observatories; 
next-generation detectors  will thus provide a key infrastructure to probe the physics of compact objects.



\end{document}